# Anodic Aqueous Electrophoretic Deposition of Titanium Dioxide Using Carboxylic Acids as Dispersing Agents

*Dorian Hanaor,[a,*] Marco Michelazzi[b], Paolo Veronesi[b], Cristina Leonelli[b], Marcello Romagnoli[b], Charles Sorrell[a]*

a: University of New South Wales, School of Materials Science and Engineering, Kensington NSW 2052, Australia
b: University of Modena and Reggio Emilia, Department of Materials and Environmental Engineering, 41100 Modena, Italy
* email: dorian@unsw.edu.au

## Abstract

The dispersion of anatase phase $TiO_2$ powder in aqueous suspensions was investigated by zeta-potential and agglomerate size analysis. The iso-electric point (IEP) of anatase was determined to be at pH 2.8 using monoprotic acids for pH adjustment. In comparison, it was found that the use of carboxylic acids, citric and oxalic, caused a decrease in zeta-potential through the adsorption of negatively charged groups to the particle surfaces. The use of these reagents was shown to enable effective anodic electrophoretic deposition (EPD) of $TiO_2$ onto graphite substrates at low pH levels with a decreased level of bubble damage in comparison with anodic EPD from basic suspensions. The results obtained demonstrate that the IEP of $TiO_2$ varies with the type of reagent used for pH adjustment. The low pH level of the IEP and the ability to decrease the zeta-potential through the use of carboxylic acids suggest that the anodic EPD of anatase is more readily facilitated than cathodic EPD.

# 1. Introduction

Titanium dioxide is distinct in its ability to function as a photocatalyst due to the particular levels of its valence and conduction bands[1]. Titanium dioxide photocatalysts are of great interest owing to their ability to facilitate various reactions of environmental benefit. In recent years applications for which $TiO_2$ photocatalysis has been investigated include:

- Generation of hydrogen [2-6]
- Dye sensitised solar cells [7-9]
- Destruction of bacteria [10-14]
- Removal of contaminants from water [15-19]
- Self-cleaning coatings [20-23]

Titanium dioxide photocatalysts have been used in various forms including powders, thin films and thick films. Although powders often show higher performance than immobilised films, the significant disadvantage in using powders is the associated difficulty in catalyst recovery [24]. It is for this reason that the immobilisation of $TiO_2$ is often carried out. Immobilised $TiO_2$ is often synthesised by the use of precursor chemicals such as titanium alkoxides or titanium tetrachloride or through the use of pre synthesised $TiO_2$ powders.

Electrophoretic deposition, EPD, is a useful technique to immobilise powders from suspensions. In the process of EPD, charged particles move towards an oppositely charged electrode and form a stable deposit. EPD is generally followed by a densification process through thermal treatment. The process of EPD has been used with suspensions of $TiO_2$ in the past [25-27]. EPD synthesis of materials has several advantages:

- Cost effectiveness
- Ability to utilise fine powders
- Homogeneity of resultant coatings

- Ability to utilise suspensions of low solids loadings
- Simple apparatus requirements
- Binder-free process

EPD has been used to fabricate thick films on a variety of conductive and non-conductive substrates as well as being used to synthesise free-standing objects. The migration and deposition of suspended particles on to a positive electrode (anode) is known as anodic deposition while EPD on to a negative electrode (cathode) is known as cathodic deposition.

Charged particles in suspension are generally surrounded by an increased concentration of ions of opposite charge. During the process of electrophoresis, a layer of these ions migrate along with the particle. The potential at the slipping plane, between the layer of counter-ions which moves along with the particle and the bulk liquid, is known as the zeta ($\zeta$) potential. Negative zeta potentials are used for anodic depositions while positive zeta potential values are necessary for cathodic EPD. Zeta potentials near zero give rise to agglomeration which is detrimental for either type of EPD.

The kinetics of electrophoretic deposition have been studied and various formulae are used in the analysis of electrophoretic processes. The Hamaker equation (Eq. 1) is a widely used kinetic model for EPD in planar geometries.

$$m = C_s \mu S E t \qquad (1)$$

Here $C_s$ is the solids loading (g cm$^{-3}$), $\mu$ is the electrophoretic mobility (cm$^2$ s$^{-1}$ V$^{-1}$), S is the deposition area (cm$^2$), E is the electric field (V cm$^{-1}$) and t is time (s) [28-31].

The electrophoretic mobility represents the velocity at which a particle moves under the influence of an applied field and is generally expressed as shown in equation 2.

$$\mu = v/E \qquad (2)$$

The velocity at which a particle moves is determined by the zeta potential. Equation 3 shows how the zeta potential can be used to express the electrophoretic mobility [29].

$$\mu = \frac{2\varepsilon_0 \varepsilon_r \zeta}{3\eta} f(\kappa r) \quad (3)$$

Here $\varepsilon_0$ is the permittivity of free space, $\varepsilon_r$ and $\eta$ are respectively the permittivity and viscosity of the suspension medium, $\zeta$ is the zeta potential of particles in suspension and $f(\kappa r)$ is the Henry coefficient, which relates the thickness of the double layer to the radius of the suspended particle.

For a situation where the double layer is thin in comparison with the particle size, this can be approximated as equation 4 [28, 32]:

$$\mu = \frac{\varepsilon_0 \varepsilon_r \zeta}{\eta} \quad (4)$$

As the Hamaker equation assumes 100% adhesion, i.e. all particles reaching the electrode remain in the EPD-formed deposit, it is appropriate to add an adhesion factor *a*, which accounts for the fraction of the deposit which remains on the electrode subsequent to extraction from the liquid medium. Therefore, if we assume a constant solids loading and electric field strength in the suspension, the Hamaker equation can be written as shown in equation 5.

$$m = \frac{a C_s \varepsilon_0 \varepsilon_r \zeta S E t}{\eta} \quad (5)$$

It can be seen from equations 4 and 5, that for a given experimental setup, suspension medium and suspended powder, variation of the zeta potential can be used to control the electrophoretic mobility and thus the EPD rate. This can be achieved by acidity regulation. Typically, the zeta potential increases with increasing pH, and pH modification can be used to control the performance of EPD processes. Zeta potential behaviour is generally consistent regardless of the acids or bases used to modify the pH, and there exists a specific pH at which the zeta potential equals zero. This pH level is widely known as the iso-electric point (IEP), or point of zero charge.

The use of citric acid and other carboxylic acids has been reported to give rise to a lower zeta potential, bring the IEP down to a lower pH, and enhance dispersion of aluminia particles in suspension [33, 34] this behaviour can be explained by the adsorption of negatively charged groups on the particle surface.

The IEP of titanium dioxide has been reported to be around pH 6 [35-37]. This would imply that suspensions most suitable for electrophoretic depositions are on either side of this value. Several studies have investigated cathodic deposition of $TiO_2$ onto conductive substrates [25, 38, 39]. And the use of basic pH levels to facilitate anodic EPD has also been reported [40] Many studies into electrophoretic deposition have used non-aqueous suspension media, typically organic media such as alcohols or acetone [31]. Water is problematic as a suspension medium due to the parasitic process of water electrolysis which takes place during the deposition and can cause gas bubbles to accumulate at the electrode surfaces to the detriment of the electrophoretically deposited coating. Despite this phenomenon, using water as a suspension medium is attractive as it has a lower environmental impact than organic media and is easier to upscale to an industrial size process.

## 2. Materials and Methods

High purity anatase powder (>99%) supplied by Merck Chemicals was used in all experiments in this work. The powder was washed with distilled water and recovered by centrifugation to remove surface contamination that may have imparted a surface charge. Suspension parameters of agglomerate size, zeta potential and electrophoretic mobility were measured using a Malvern Instruments Nano Series Zetasizer. All suspensions were made using distilled water as a suspension medium. Due to the high opacity of anatase suspensions, for zeta potential and agglomerate size analysis suspensions of 0.05 wt% ($5 \times 10^{-4}$ g cm$^{-3}$) were used. To determine the typical effect of acidity on zeta potential and agglomeration, pH levels were varied with nitric acid / ammonium hydroxide and hydrochloric acid / sodium hydroxide. The effects of carboxylic acids on suspension properties were investigated by the use of citric and oxalic acids as pH varying agents. To achieve basic pH levels, solutions of the carboxylic acids with NaOH at a 1:4 molar ratio were employed to give basic pH levels while maintaining levels of carboxylic groups in solution sufficient to saturate particles surfaces [33].

Anodic electrophoretic depositions were carried out onto 25 x 25 x 2 mm graphite substrates immersed in the suspensions to a depth of 10 mm. Prior to depositions, the graphite substrates were ultrasonically cleaned in acetone of purity >99.5%, from Sigma Aldrich (<0.5% $H_2O$, <0.05% isopropanol, <0.05% methanol, <0.001% evaporation residue). Subsequent to acetone cleaning, substrates were dried at 110 °C and adhesive tape was applied as an insulating backing. It is possible that residues resulting from acetone cleaning have a detrimental effect on the adhesion of thick films to the graphite substrates, however as the evaporation residue of the acetone used is reported at <0.001% this is unlikely to be a significant factor in the current work. Graphite substrates were chosen for the EPD of $TiO_2$ as carbon diffusion into $TiO_2$ coatings may improve photocatalytic performance as suggested by results of other work [41]. Furthermore the graphite may act as a reducing agent and increase oxygen vacancy levels in the anatase thus lowering the anatase to rutile transformation temperature and enabling the formation of mixed anatase/rutile $TiO_2$ photocatalysts at temperatures lower than what would otherwise be possible on metallic substrates [42]. The formation of such a mixed- phase composition may too be favourable for photocatalytic performance based on previous publications [43-45].

Anodic EPD experiments used aqueous suspensions of 1 wt% (0.01 g $cm^{-3}$) anatase solids loadings. Depositions utilised the same pH adjustment reagents used for zeta potential and agglomerate size analysis. Electrophoretic depositions lasting 10 minutes were carried out using a 10 V DC power supply with a strip of gold foil as a cathode (counter electrode) and the graphite substrates as anodes (working electrodes). These electrodes were separated by 20 mm. Electrical contact to the electrodes was made with alligator clips. The applied voltage was maintained during the slow extraction of the working electrode from the suspension to minimise coating removal during extraction.

The quality of the coatings achieved by EPD was assessed by optical microscopy and scanning electron microscopy (SEM) using a Hitachi S3400 microscope.

## 3. Results

### *3.1. Suspension properties*

Standard variation of zeta potential was investigated by varying pH levels with commonly used monoprotic acids and bases. This is shown in figure 1. The resultant variation of zeta potential with acidity appears consistent for both sets of reagents and it can be seen that the IEP appears to be around pH 2.80. This figure is lower than what has been reported previously as the IEP of $TiO_2$. Values for electrophoretic mobility were also recorded from the apparatus used. These values were found to show a linear relationship with the values recorded for zeta potential. This is expected as parameters in equation 4 remain constant apart from $\zeta$ potential.

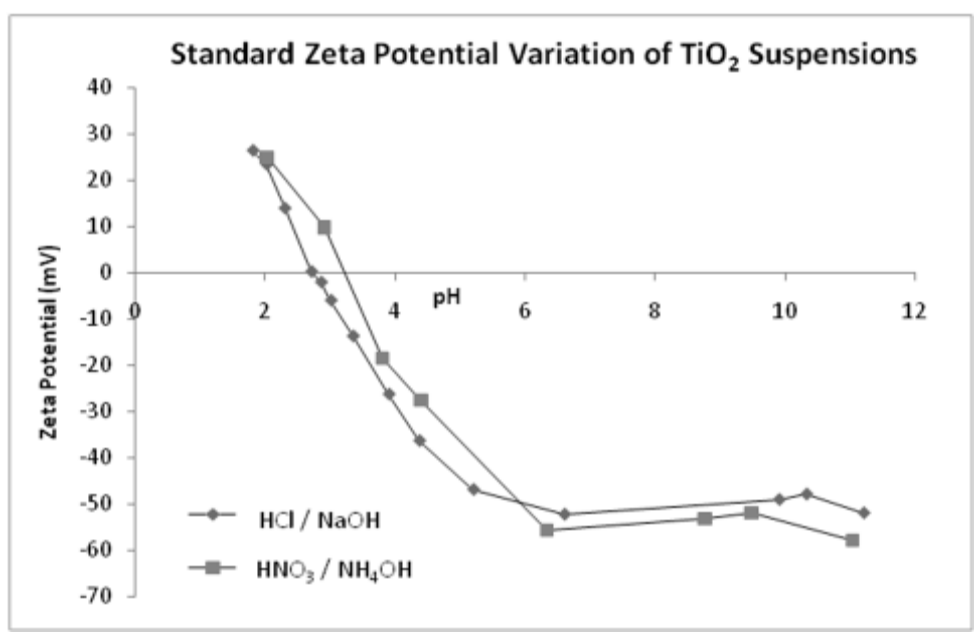

**Figure 1. Zeta potential of $TiO_2$ as a function of pH value adjusted with typical acids and bases**

Figure 2 shows the typical agglomeration behaviour of $TiO_2$ suspensions using nitric acid and ammonium hydroxide as pH adjusting agents. Due to equipment limitations non-agglomerated suspensions were found generally to return a value of 300-400nm for agglomerate size. This is not necessarily a true indication of particle size in non-agglomerated suspensions.

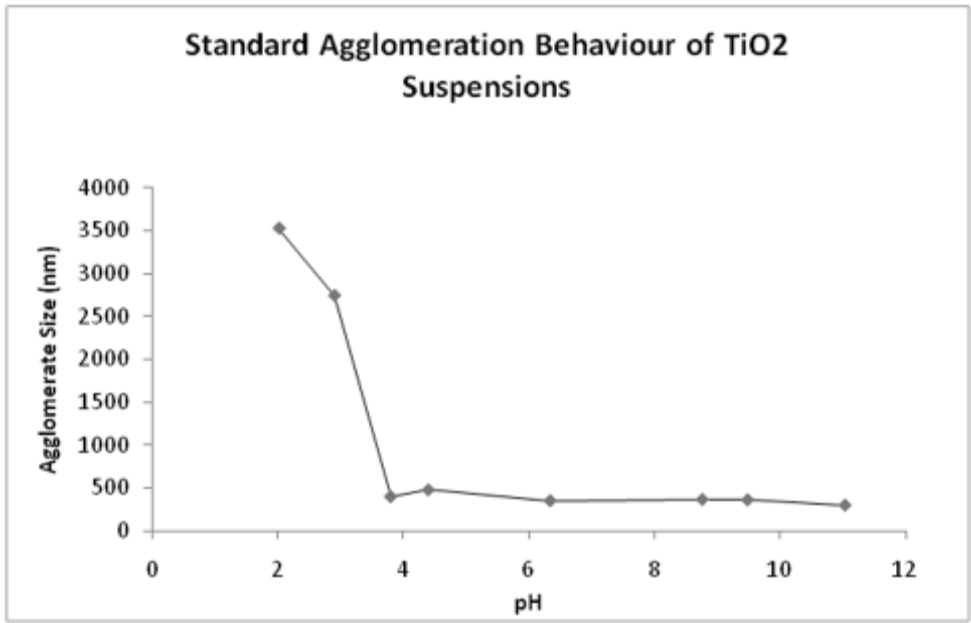

**Figure 2. Agglomerate size of TiO$_2$ suspensions as a function of pH value adjusted with nitric acid and ammonium hydroxide**

In contrast to the use of monoprotic acids, the use of citric acid resulted in a negative zeta potential at all pH levels. The zeta potential rises as pH decreases however when using citric acid for pH adjustment, the suspension did not appear to reach a point of zero charge (IEP). Flocculation was observed to take place at lower pH levels despite the negative zeta potential values recorded.

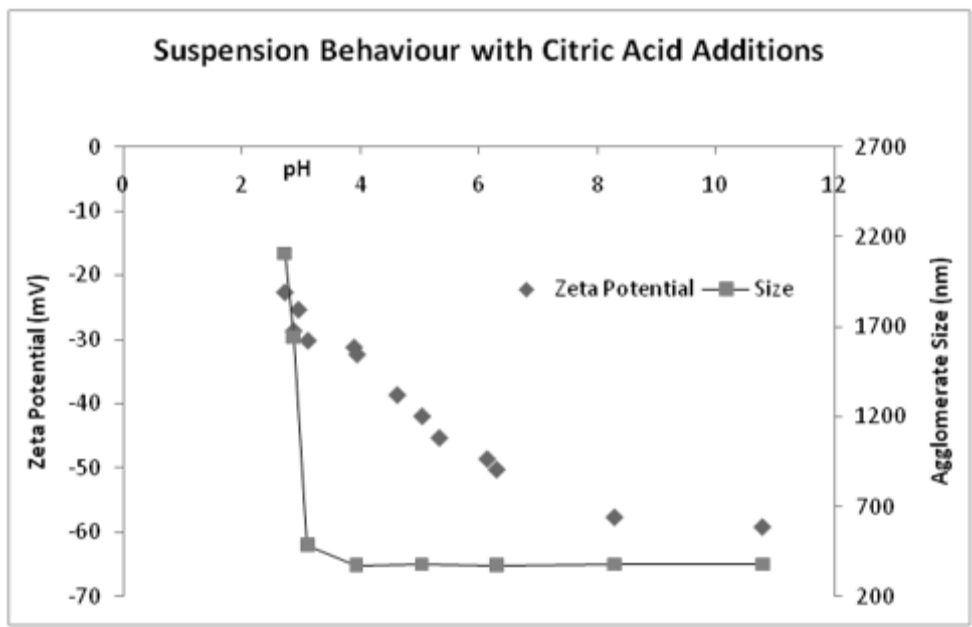

**Figure 3. Zeta potential and agglomerate size as a function of pH value adjusted with citric acid**

The use of oxalic acid resulted in negative zeta potential values slightly lower than those obtained with the use of citric acid. Again a point of zero charge was not reached. In similarity to the case of citric acid, flocculation occurred in suspensions of low pH showing negative zeta potentials.

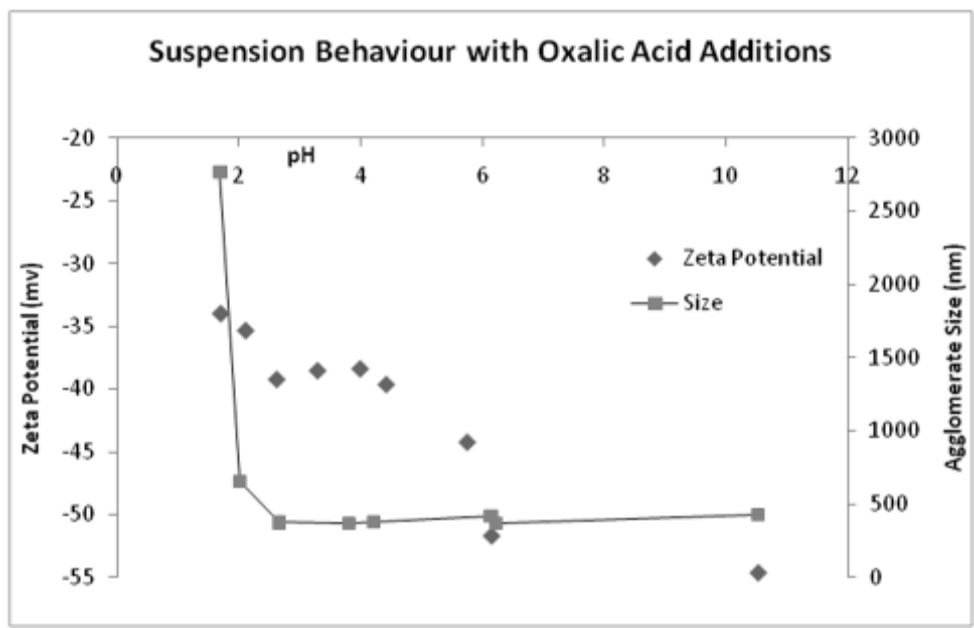

**Figure 4. Zeta potential and agglomerate size as a function of pH value adjusted with oxalic acid**

### *3.2. EPD Coatings*

Electrophoretic depositions were carried out using the same reagents described in the preceding section to impart acidic and basic pH levels to the suspensions. The weight of the graphite substrate before and after the deposition along with the dimensions of the coated area was used to determine the deposited mass per unit area, reported in mg cm$^{-2}$. The calculated mass was determined using the Hamaker equation (eq. 1) in conjunction with electrophoretic mobility readings. This calculation did not take into consideration the insulation effects of the coating on the electrical field strength, which was taken as 5 V cm$^{-1}$, and the decrease in the solids loading of the suspension as deposition proceeded. The results of electrophoretic depositions are outlined in table 1.

Table 1. Electrophoretic deposition data

| | Reagent | pH | ζ potential (mv) | µ ($cm^2V^{-1}S^{-1}$) $x10^{-4}$ | Deposited Mass (mg $cm^{-2}$) | Calculated Mass (mg $cm^{-2}$) |
|---|---|---|---|---|---|---|
| Acidic | Nitric Acid | 3.62 | -10.86 | -0.85 | 0.88 | 2.54 |
| | Citric Acid | 3.80 | -33.73 | -2.63 | 4.14 | 7.89 |
| | Oxalic Acid | 3.30 | -38.73 | -3.02 | 7.82 | 9.06 |
| Basic | NaOH | 10.71 | -59.88 | -4.67 | 8.18 | 14.01 |
| | Citric + NaOH | 10.57 | -59.28 | -4.66 | 8.52 | 14.00 |
| | Oxalic + NaOH | 11.80 | -56.92 | -4.43 | 12.36 | 13.32 |

The comparison of the calculated deposit mass according to eq. 1 with the measured deposit mass resulting from anodic electrophoretic depositions from acidic suspensions using different pH adjustment agents is illustrated in figure 5. The use of carboxylic acids significantly raised the deposition rate at lower pH levels. At basic pH levels the use of carboxylic acids had a less marked effect on deposition rates although some effect was observed, this can be seen in figure 6.

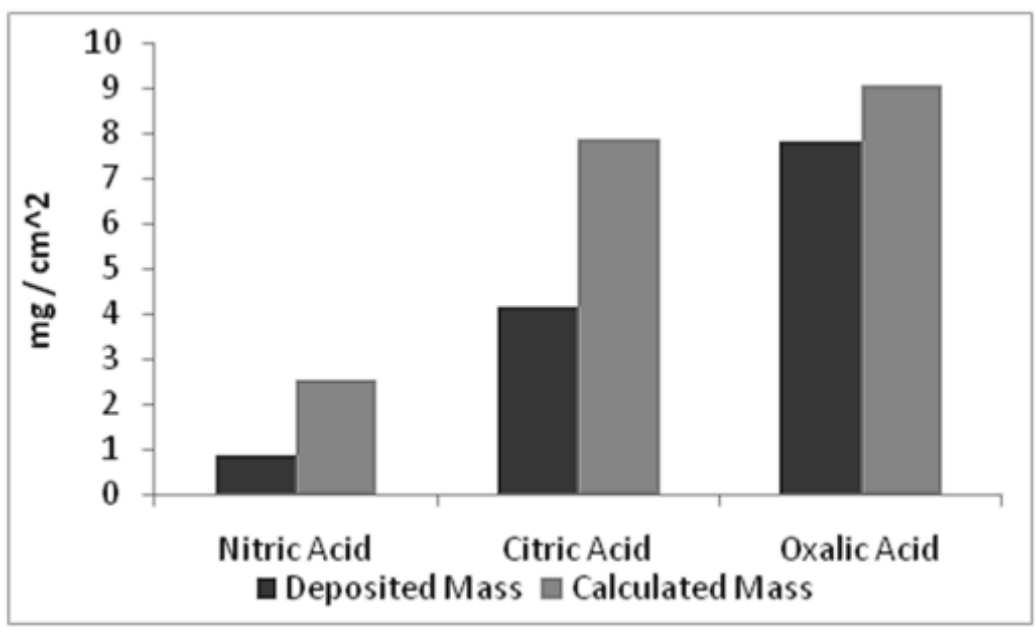

Figure 5. Comparison of actual deposit mass with calculated deposit mass from acidic suspensions using different pH adjusting agents

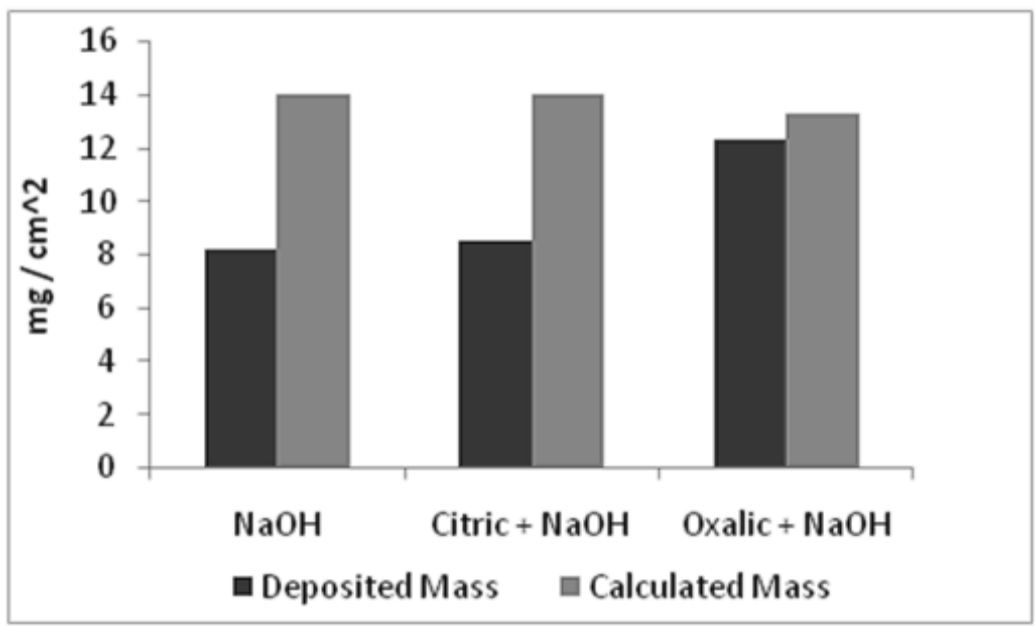

**Figure 6. Comparison of actual deposit mass with calculated deposit mass from basic suspensions using different pH adjusting agents**

Images of the deposited material were taken by optical microscopy and scanning electron microscopy. These images show surfaces marked with pinholes resulting from gas bubbles trapped in the coating. These gas bubbles are likely to be the result of the electrolysis of water, a parasitic process discussed earlier. Coatings made with acidic pH levels adjusted by oxalic and citric acids were fairly consistent with many small pinholes of 5-10 µm in size (Figure 7). Coatings made from acidic suspensions with nitric acid were irregular and coverage was poor (Figure 8). This is consistent with the low level of deposited mass per unit area measured. Coatings made from basic suspensions showed more extensive evidence of gas bubble damage to the electrophoretically deposited coating, with holes or craters ranging from 20-50 µm in size as seen in Figure 9, suggesting greater levels of water electrolysis or larger gas bubbles at the electrode / suspension interface.

The microstructure of the deposit can be seen in the SEM micrograph in Figure 10. The grains are in the region of 200 nm in size. This is due to the morphology of the anatase powder used in the EPD processes this structure was consistent across all samples fabricated.

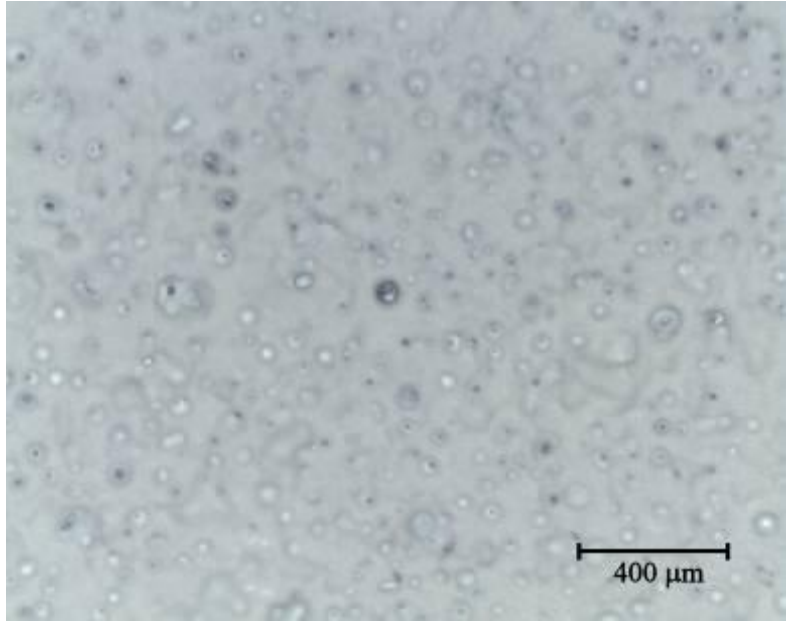

**Figure 7. EPD coating typical of those produced from acidic suspensions with pH adjustment by carboxylic acids**

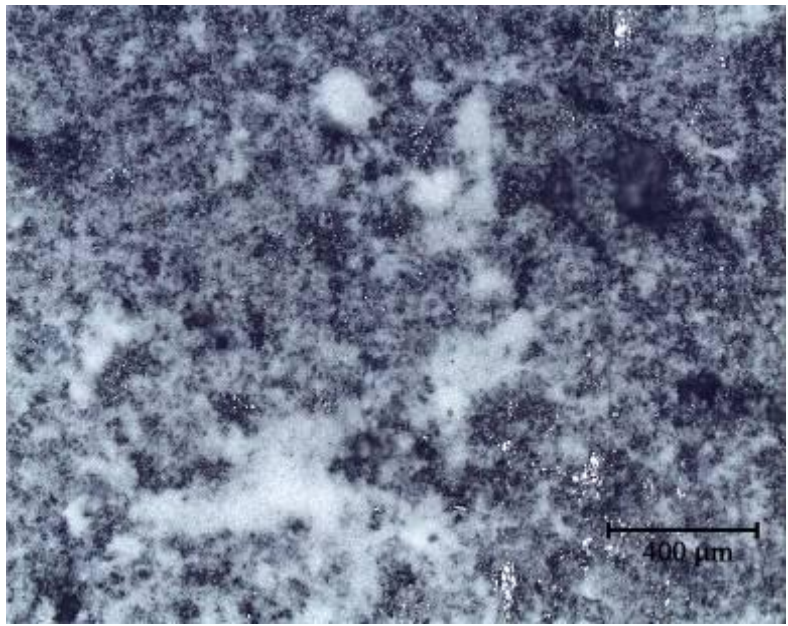

**Figure 8. EPD coating produced from an acidic suspension with pH adjustment by Nitric Acid**

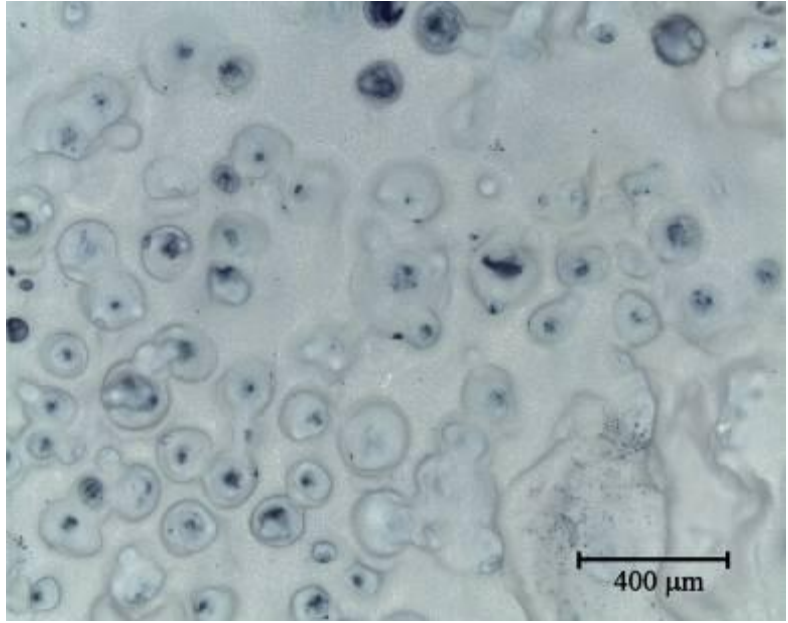

**Figure 9. EPD coating typical of those produced from basic suspensions with pH adjustment by sodium hydroxide**

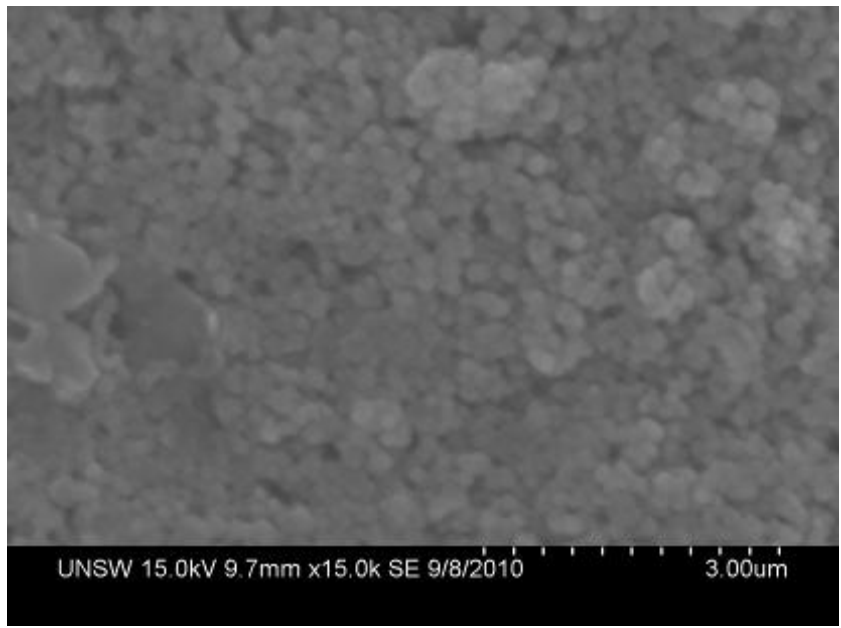

**Figure 10. Microstructure of deposited anatase**

## 4. Discussion

The variation of the zeta potential of aqueous $TiO_2$ suspensions with conventional pH adjustment showed typical behaviour with positive values at low pH values decreasing to negative values at higher pH levels. The behaviour of such suspensions appeared to be divergent from what has been reported previously in that the IEP was determined to at pH 2.8. This is a lower figure than what has been observed in other work and suggests that anodic EPD is more readily facilitated than cathodic EPD with aqueous suspensions of $TiO_2$. The discrepancy between IEP valued from this work and those obtained elsewhere could be due to impurities in the material used as low levels of impurities in the suspended particles or in the suspension medium can profoundly affect the zeta potential [46, 47]. Although the anatase used in this work was of high purity and had undergone washing, it is possible that the presence of acidic groups on the surface brought the IEP down to lower levels as reported elsewhere [48, 49].

In comparison with pH adjustment by monoprotic acids, hydrochloric and nitric, the use of citric and oxalic acids, tri-carboxylic and bi-carboxylic acids respectively, was shown to bring about significantly lower zeta potential values and thus lower electrophoretic mobility values in suspensions of $TiO_2$ across all pH levels these values remained negative even at low pH levels. This phenomenon is likely to be due to negative citrate and oxalate ions adsorbed on the $TiO_2$ surfaces and imparting a negative charge to these particles. It has been reported that negative monovalent citrate ions show stronger adsorption to surfaces in comparison with the fully protonated citric acid and this preferential adsorption may occur with the use of other carboxylic acids [50]. This preferential adsorption of negatively charged groups may explain the negative zeta potential values imparted by the carboxylic acids used in this work at lower pH values. Similar observations of lowered zeta potential values were reported from experiments using citric acid as a low molecular weight dispersant for $Al_2O_3$ particles in aqueous suspension where it was reported that two of the three carboxylic groups of the citric acid coordinate to the alumina surface [33].

Despite the negative zeta potential values, acidic $TiO_2$ suspensions adjusted by means of carboxylic acids used in this work exhibited agglomeration at pH levels around 2. As citric and oxalic are weak acids, high concentrations of these reagents were necessary to achieve low pH levels. It is likely that the high levels of protonated citric and oxalic groups in aqueous solution lead to a decrease in the volume of the liquid sphere which moves along with the particle in suspension and thus allows the particles to approach each other, facilitating agglomeration [27, 32, 33].

Citric and oxalic acids were successfully used to facilitate anodic electrophoretic depositions of $TiO_2$ onto graphite substrates at low pH levels. The negative zeta potential achieved through the use of these reagents enabled high deposition rates in comparison with depositions which utilised monoprotic nitric acid to impart acidity. While the measurement of the deposited mass per unit area is likely to vary due to experimental inaccuracies associated with weighing the substrates before and after deposition, it was clear that the use of carboxylic acids to impart low pH levels was advantageous for anodic EPD from acidic suspensions. Anodic EPD from basic suspensions were fairly rapid with and without the use of carboxylic acids. Although the use of carboxylic acids along in basic suspensions did seem to bring about a higher deposit mass, this improvement was less significant than in the case of acidic suspensions. The good levels of deposit mass in EPD from all basic suspensions are likely to be a result of low zeta potential levels associated with high pH suspensions, however the apparent drawback of such depositions was the apparent increased extent of water electrolysis that was evident through the presence of large craters due to gas bubbles in the deposited thick films achieved from basic suspensions. This phenomenon can be explained by the increased electrolysis of water at high pH levels that has been reported elsewhere [51]. This highlights the advantages of the use of lower pH suspensions for aqueous electrophoretic depositions and the use of carboxylic acids as low molecular weight dispersants to achieve such depositions.

The use of polyelectrolyte dispersants such as poly-acrylates is a widespread method to enhance the dispersion of ceramic particles in suspension [52, 53]. The use of carboxylic acids as alternative lower molecular weight dispersants has advantages over the use of long-chain molecules of due to higher adsorption ability, greater chemical stability, lower cost and a lower environmental impact than such high molecular weight additives [34, 54, 55].

Further work may investigate the sintering of $TiO_2$ coatings such as those synthesised in this work and the resultant photocatalytic performance of these materials.

## 5. Conclusions

Citric and oxalic acids, compounds with multiple carboxylic groups bind to $TiO_2$ particle surfaces, and impart strongly negative zeta potential values and a greater electrophoretic mobility to these particles in aqueous suspensions. Thus such reagents can be used as low molecular weight dispersants for aqueous suspensions of $TiO_2$

Effective anodic electrophoretic deposition from acidic suspensions can be facilitated through the use of carboxylic acid additions. This may improve the quality of the electrophoretically deposited coating in comparison with the use of basic suspensions through lower levels of water electrolysis and associated gas bubbles in the deposits.